\documentclass[cameraready]{Interspeech}
\title{Towards Unified Song Generation and Singing Voice Conversion with
Accompaniment Co-Generation}

\author[affiliation={1}]{Ziyu}{Zhang}
\author[affiliation={2}]{Chunyu}{Qiang}
\author[affiliation={3}]{Xiaopeng}{Wang}
\author[affiliation={4}]{Yuxin}{Guo}
\author[affiliation={2}]{Kang}{Yin}
\author[affiliation={1}]{Wenjie}{Tian}
\author[affiliation={1}]{Jingbin}{Hu}
\author[affiliation={1}]{Tianlun}{Zuo}
\author[affiliation={1}]{Zhao}{Guo}
\author[affiliation={2}]{Teng}{Ma}
\author[affiliation={2}]{Yuzhe}{Liang}
\author[affiliation={2}]{Chen}{Zhang}
\author[affiliation={1}, correspondingauthor]{Lei}{Xie}

\address{
    $^1$ ASLP@NPU, Northwestern Polytechnical University, China \\
    $^2$ Kuaishou Technology, China   
    $^3$ Beijing Institute of Technology, China \\
    $^4$ Institute of Automation, Chinese Academy of Sciences, China 
}

\email{ziyu\_zhang@mail.nwpu.edu.cn, lxie@nwpu.edu.cn}

\usepackage{comment}
\usepackage{multirow}

\usepackage{latexsym}
\usepackage{hyperref}
\usepackage{enumitem}
\usepackage{booktabs} % 用于绘制专业的三线表
\usepackage{amsmath}  % 用于数学符号,如 \uparrow
\usepackage{amssymb}
\usepackage{subcaption}
\usepackage{graphicx}
\usepackage[T1]{fontenc}
\usepackage[utf8]{inputenc}
\usepackage{microtype}
\usepackage{inconsolata}
\begin{document}

\maketitle

\vspace{-3ex}
\begin{abstract}
    While song generation and singing voice conversion (SVC) have evolved significantly, they have long been developed isolated: the former lacks zero-shot speaker cloning, while the latter overlooks vocal-accompaniment synergy. To bridge this gap, we propose UniSinger, the first end-to-end framework unifying speaker cloning song generation and accompaniment co-generation SVC. Building on the multimodal diffusion transformer, we construct a unified speaker embedding space transferring speaker representation from SVC to song generation, endowing fine-grained cross-task timbre control. To mitigate multi-task optimization conflicts, we design a curriculum learning strategy using task-specific modality masking to guide the model to gradually master the generative mechanisms among semantic content, vocal timbre, and accompaniment. Experiments show state-of-the-art performance on both tasks and realizes complementary benefits, offering new possibilities for intelligent music production.
    
\end{abstract}  

\vspace{-1ex}
\section{Introduction}
% \vspace{-1ex}

% 第一段,点出背景
Recent breakthroughs in generative AI have driven the evolution of song generation and SVC. 
%统一建模这两个任务对于音乐创作任务来说非常重要非常好。
Unified modeling of these two tasks holds significant importance for advancing the field of music creation. 
% 然而这两个任务各自都存在问题。
However, existing methods for both independent tasks still face significant limitations.
%具体来说,歌曲生成：
Specifically, for song generation, while pioneering works~\cite{dhariwal2020jukebox,van2017neural} and large-scale models~\cite{copet2023simple,agostinelli2023musiclm} established robust text-to-song generation, they lack fine-grained vocal control. Subsequent models~\cite{yang2025songeditor, ding2024songcomposer, ning2025diffrhythm} improved generation quality but still struggle with speaker cloning. Although emerging methods~\cite{chen2025diffrhythm+, yuan2025yue} have achieved zero-shot voice cloning, they continue to treat song generation and SVC as isolated problems.
% 对于SVC：
Conversely, regarding SVC, while diffusion-based SVC models~\cite{liu2021diffsvc, bai2025hq} have achieved high-fidelity vocal reconstruction, and recent advances~\cite{sha2024neural, lu2024comosvc} have further improved inference efficiency, these methods typically neglect the acoustic synergy between vocals and background music (BGM). 
%Although cascaded approach~\cite{donahue2023singsong} try to generate accompaniment for vocals, it struggles to ensure emotional and rhythmic consistency. 
% 承接下一段 实现任务统一面临的阻碍
However, achieving the unified modeling of these two tasks requires not only addressing issues of individual tasks but also resolving the fundamental conflicts arising from the deep integration of two heterogeneous tasks.

% 第二段,点出统一建模的困境
Specifically, unifying these two tasks is hindered by critical obstacles at both the model and training paradigm levels. 
% 模型层面
At the model level, SVC is an Audio-to-Audio mapping task requiring acoustic inputs, while song generation is a Text-to-Audio task driven by textual inputs. Therefore, mapping heterogeneous inputs to a shared latent space within a unified architecture is difficult. Moreover, establishing a unified cross-task speaker injection mechanism proves highly challenging.
% 训练框架层面
Regarding the training paradigm, SVC aims to achieve vocal timbre disentanglement, whereas song generation focuses on creating melodies from scratch. These differences create a dilemma for joint training. Naive multi-task integration often leads to gradient conflicts and local optima where model either neglects fine-grained vocal details or yields incoherent musical structures.

% 第三段,点出我们的策略和贡献
% \vspace{-3.2ex}
To address these challenges, we propose \textbf{UniSinger}, the first framework unifying speaker cloning song generation and accompaniment co-generation SVC. At the model level, we employ a multi-modal input module to project diverse inputs into a shared latent space, leveraging an MM-DiT backbone for deep representation alignment. For speaker injection, we establish a cross-task  speaker embedding space to ensure fine-grained vocal control. Finally, at the training paradigm level, we design a progressive curriculum learning strategy with task-specific modality masking. By selectively masking irrelevant modalities, this approach progressively guides the model to master both reference-driven vocal synthesis and text-driven accompaniment generation, thereby enhancing cross-modal alignment.
% 在此基础上,我们还发现
Building on this architecture, we further identify a generalizable principle: the two tasks serve as mutual priors. Song generation provides global musical structure to refine SVC's prosody and harmony, while SVC's robust semantic modeling reduces pronunciation errors in song generation. Audio samples are available  here~\footnote{~\url{https://ziyu6.github.io/UniSinger/}}. Our main contributions are summarized as follows:

% \vspace{-3.2ex}
\begin{itemize}[nosep]
    \item We unify Song Generation and SVC for the first time, enabling fine-grained vocal control and accompaniment co-generation.
    \item We propose an MM-DiT architecture with a unified speaker embedding space to achieve multi-modal alignment and cross-task timbre transfer. 
    \item We design a progressive curriculum learning strategy with task-specific modality masking to resolve gradient conflicts and achieve cross-modal alignment.
    \item Experiments demonstrate state-of-the-art performance and identify generalizable inter-task mutual benefits.
\end{itemize}

\begin{figure*}
    \centering
    \includegraphics[width=\linewidth]{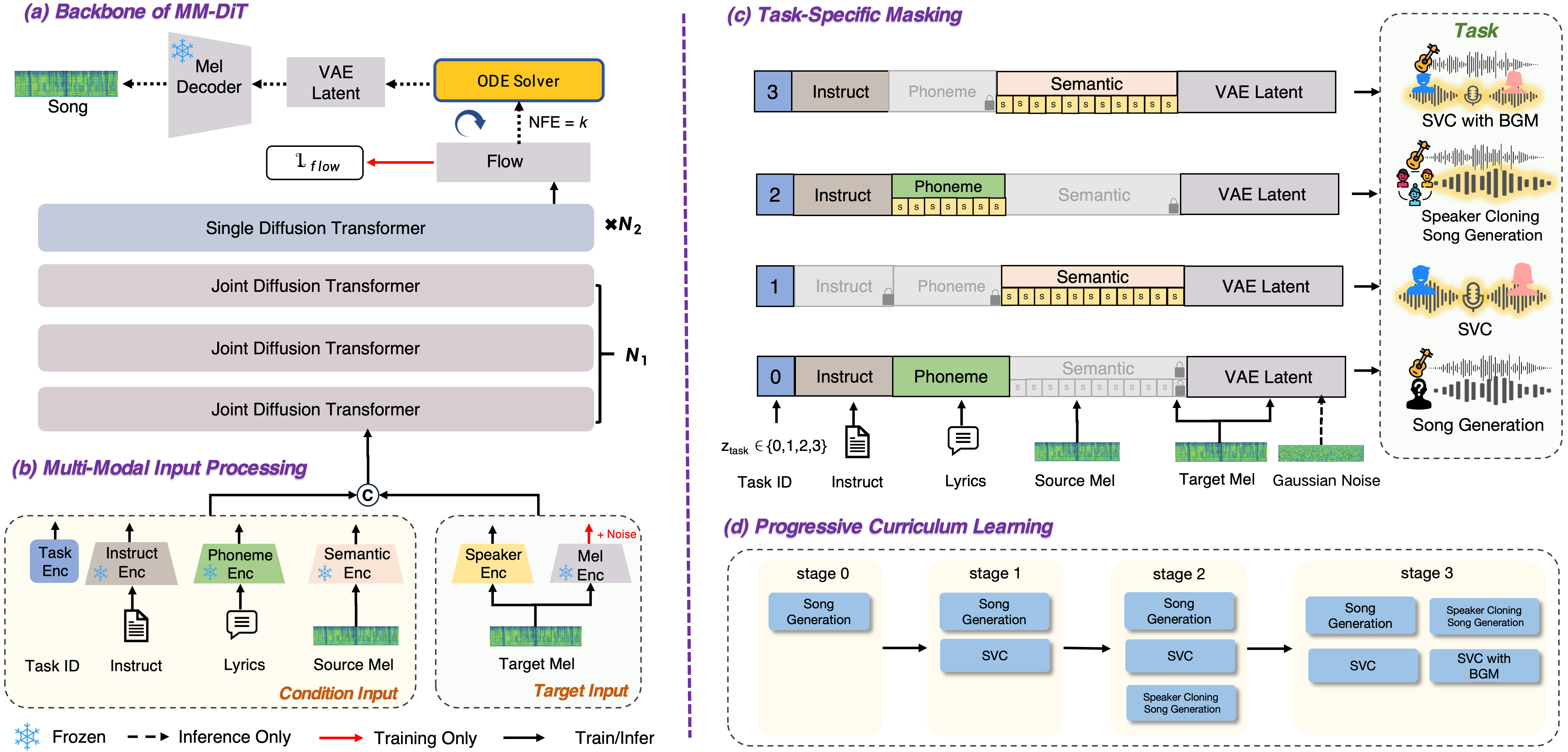}
    \caption{The overall architecture of UniSinger. The left panel shows the MM-DiT backbone and multi-modal input processing module. The right panel details the progressive curriculum learning strategy with task-specific modality masking.}
    \vspace{-3ex}
    \label{fig:placeholder}
\end{figure*}
% \vspace{-5ex}

\vspace{-2ex}
\section{Method}
\vspace{-1ex}

\subsection{Overview}
\vspace{-1ex}

% As illustrated in Figure, UniSinger is designed as a unified generative framework based on Flow Matching. The overall architecture enables the joint modeling through three integral components.

% First, to bridge the gap between heterogeneous tasks, we employ a \textbf{Multi-Modal Input Processing} module. This module projects diverse conditions—including text instructions, phonemes and audios—into a shared feature space, enabling flexible conditioning. Second, to mitigate optimization conflicts and facilitate cross-task synergy, we introduce a \textbf{Progressive Curriculum Learning} strategy. By applying \textbf{task-specific modality masks}, we guide the model through four stages, transitioning from fundamental singing synthesis to complex accompaniment-coordinated generation. Finally, the core generative process is powered by a Multimodal Diffusion Transformer (MM-DiT) Backbone, which models the conditional distribution of audio latents via Flow Matching.

% As illustrated in Figure~\ref{fig:placeholder}, UniSinger is a unified generative framework comprising three modules. First, Multi-Modal Input Processing projects diverse conditions into a shared latent space to enable flexible conditioning. Second, a Progressive Curriculum Learning strategy employs task-specific modality masks to guide the model from foundational synthesis to complex accompaniment-coordinated generation. Finally, the core generative process is driven by a MM-DiT backbone based on flow matching, which models the conditional distribution of audio latents.

As shown in Figure~\ref{fig:placeholder}, UniSinger consists of four core components: (1) multi-modal input processing, which projects diverse conditions into a shared latent space; (2) progressive curriculum learning, employing task-specific modality masks to guide generation from foundational synthesis to complex accompaniment coordination ; (3) cross-task speaker embedding space, which ensuring consistent vocal identity preservation across tasks; and (4) an MM-DiT backbone, which models audio latent distributions via flow matching.

\vspace{-1ex}
\subsection{Multi-Modal Input Processing}
\vspace{-1ex}

% In this section, we demonstrate the processing pipeline that transforms raw multi-modal inputs into a unified joint representation. 
% As shown in Figure~\ref{fig:placeholder}(a), we employ a suite of pre-trained encoders to project multi-modal inputs into a shared latent space: (1) Instruct \& Phonetic Encoder: Frozen Qwen2.5-7B~\cite{hui2024qwen2} for text instructions and a Zipformer-based~\cite{zhu2025zipvoice} encoder for lyric phonemes; (2) Semantic Encoder: A GPT-SoVITS~\footnote{~\url{https://github.com/RVC-Boss/GPT-SoVITS}} pipeline which cascades HuBERT and Vector Quantization (VQ)~\cite{gray1984vector}  to extract speaker-independent linguistic features; (3) Speaker Encoder: CAM++~\cite{wang2023cam++} to provide a global speaker embedding for vocal identity; (4) Audio Codec: We train a VAE extending our SecoustiCodec~\cite{qiang2025secousticodec} that compresses 44.1kHz audio into a $1024\times$ downsampled latent space, serving as the DiT regression target.

In this section, we demonstrate the processing pipeline that transforms raw multi-modal inputs into a unified joint representation. As shown in Figure~\ref{fig:placeholder}(b), we employ a suite of pre-trained encoders to project multi-modal inputs into a shared latent space:

% \begin{itemize}
\noindent \textbf{Instruct \& Phonetic Encoder:} We utilize a frozen Qwen2.5-7B~\cite{hui2024qwen2} for text instructions and a Zipformer-based~\cite{zhu2025zipvoice} encoder for lyric phonemes.

\noindent \textbf{Semantic Encoder:} We adopt a So-VITS-SVC~\footnote{~\url{https://github.com/RVC-Boss/GPT-SoVITS}} pipeline that cascades HuBERT~\cite{hsu2021hubert} and Vector Quantization (VQ)~\cite{gray1984vector} to extract speaker-independent linguistic features.

\noindent \textbf{Speaker Encoder:} CAM++~\cite{wang2023cam++} is employed to provide a global speaker embedding for vocal identity.

\noindent \textbf{Audio Codec:} We train a VAE extending SecoustiCodec~\cite{qiang2025secousticodec} to compress 44.1kHz audio into a $1024\times$ downsampled latent space, serving as the DiT regression target.
% \end{itemize}

% \vspace{-1ex}
% \subsubsection{Input Encoders}
% \vspace{-1ex}

% In this section, we demonstrate the processing pipeline that transforms raw multi-modal inputs into a unified joint representation. As shown in Figure~\ref{fig:placeholder}(a), we employ a set of encoders to project various inputs into a shared latent space.

% \noindent\textbf{Instruct \& Phonetic Encoders.} 
% For text instructions, we utilize Qwen2.5-7B~\cite{hui2024qwen2} as the frozen instruct encoder. For lyrics, we employ a Zipformer-based~\cite{zhu2025zipvoice} encoder to extract phoneme embeddings.

% \noindent\textbf{Semantic Encoder.} 
% To support SVC, we extract speaker-independent linguistic features using the pre-trained semantic pipeline from GPT-SoVITS~\footnote{~\url{https://github.com/RVC-Boss/GPT-SoVITS}}, which cascades a HuBERT model with a Vector Quantization (VQ)~\cite{gray1984vector} module. 

% \noindent\textbf{Speaker Encoder.} We employ the pre-trained CAM++~\cite{wang2023cam++} model to extract a global speaker embedding. This embedding functions as a unified cross-task representation, supplying vocal identity for both SVC and song generation, Details can be found in Sec ~\ref{sec:spk}.

% \noindent\textbf{Latent Audio Codec.} 
% Instead of operating on raw waveforms, we train a Variational Autoencoder (VAE)~\cite{pinheiro2021variational} extending our SecoustiCodec~\cite{qiang2025secousticodec}. The VAE encoder compresses 44.1kHz audio into a compact latent space with 1024$\times$ downsampling, serving as the regression target for the DiT.

Before entering the backbone, embeddings are integrated into two primary streams: the condition input and the target input. The condition input is formed by temporally concatenating masked instruction, phoneme, and semantic embeddings. Global speaker embeddings are broadcast and concatenated along the feature dimension to ensure consistent vocal conditioning. For the target input, Gaussian noise is added to clean latents during training, while inference initiates directly from noise. Finally, $C_{\text{cond}}$ and $C_{\text{target}}$ are concatenated temporally for joint modeling.

\begin{table*}[t]
    \centering
    \footnotesize 
    % 表格标题
    \caption{The objective evaluation of song generation models. Best results are in \textbf{bold}, and second-best results are \underline{underlined}.}
    \vspace{-2ex}
    \label{tab:song_gen_results}
    
    % 定义列格式:l代表左对齐,c代表居中对齐
    % 改动:现在有 1(模型) + 3(PER/Spk/CLaMP) + 5(SongEval) = 9列
    \begin{tabular}{lcccccccc}
        \toprule % 顶部粗线
        
        % 表头第一行
        % 增加 CLaMP 3 列,注意它也是跨两行 (multirow)
        \multirow{2}{*}{\textbf{Model}} & \multirow{2}{*}{\textbf{PER}$\downarrow$} & \multirow{2}{*}{\textbf{Spk-Sim}$\uparrow$} & \multirow{2}{*}{\textbf{CLaMP 3}$\uparrow$} & \multicolumn{5}{c}{\textbf{SongEval}} \\ 
        
        % 改动:cmidrule 现在横跨第5列到第9列
        \cmidrule(lr){5-9} 
        
        % 表头第二行 (SongEval 的子指标)
        % 改动:现在前面要留 4 个空位 (& & & &) 给上面的 multirow
        & & & & Coh$\uparrow$ & Mem$\uparrow$ & NVBP$\uparrow$ & CSS$\uparrow$ & OM$\uparrow$\\ 
        
        \midrule % 中部横线
        
        % 数据行 (注意:现在每行需要填写 9 个数据)
        % 我在 Spk-Sim 和 SongEval 之间插入了 CLaMP 3 的占位数据
        % GT & 0.051 & 0.985 & 0.850 & 4.078 & 4.005 & 3.776 & 3.760 & 3.825 \\
        SongLM ~\cite{yang2025songeditor} & 28.32\% & 55.43\% & 0.120 & 2.538 & 2.232 & 2.614 & 2.412 & 2.212 \\
        YuE ~\cite{yuan2025yue} & 22.14\% & \underline{65.15\%} & \textbf{0.249} & 3.723 & \underline{3.681} & \underline{3.278} & \underline{3.423} & \textbf{3.458} \\ 
        ACE-Step ~\cite{gong2025ace} & 26.72\% & 52.32\% & 0.117 & 3.322 & 2.817 & 2.902 & 2.823 & 3.165 \\
        DiffRhythm+
        ~\cite{chen2025diffrhythm+}& \underline{20.72\%} & 64.21\% & 0.157 & \underline{3.750} & 3.614 & 3.202 & 3.417 & 3.287 \\

        \midrule % 中部分隔线
        
        UniSinger\_Song & \textbf{19.61\%} & \textbf{68.85\%} & \underline{0.165} & \textbf{3.768} & \textbf{3.738} & \textbf{3.312} & \textbf{3.527} & \underline{3.419} \\
        
        \bottomrule % 底部粗线
    \end{tabular}
\end{table*}
% 图1
\begin{figure*}[t]
    \centering
    % 第一张图
    \begin{subfigure}[b]{0.29\textwidth}
        \centering
        \includegraphics[width=\linewidth]{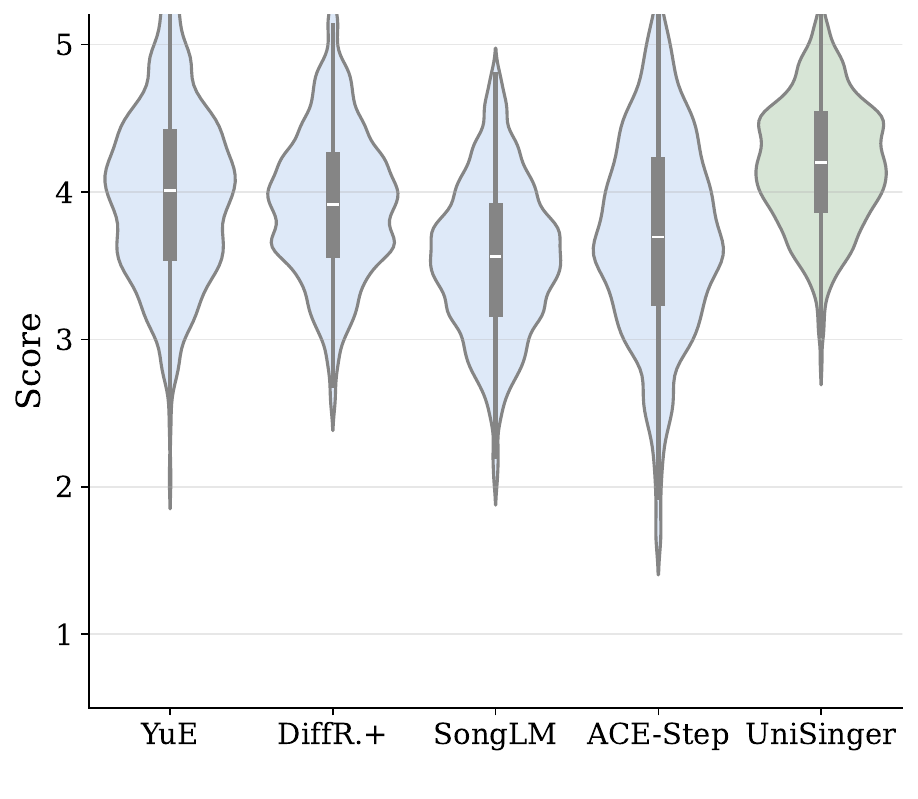}
        \vspace{-4ex}
        \caption{Intelligibility}
        \vspace{-1ex}
        \label{fig:sub1}
    \end{subfigure}% <--- 必须加 %,消除换行空格
    \hspace{1mm}% <--- 必须加 %,消除换行空格
    % 第二张图
    \begin{subfigure}[b]{0.29\textwidth}
        \centering
        \includegraphics[width=\linewidth]{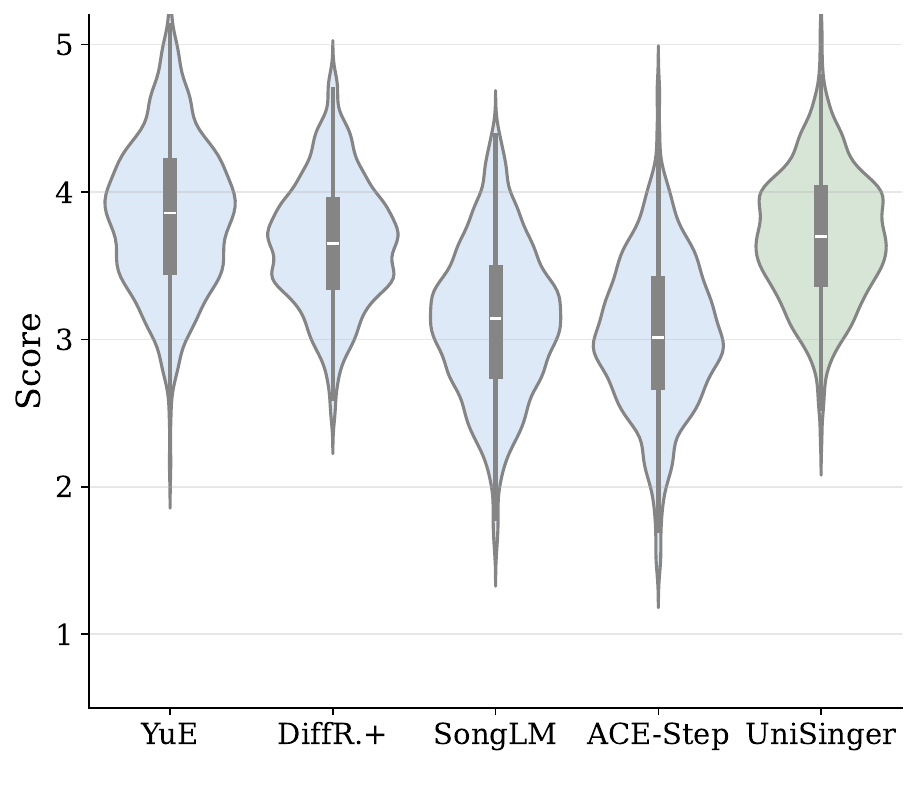}
        \vspace{-4ex}
        \caption{Similarity}
        \vspace{-1ex}
        \label{fig:sub2}
    \end{subfigure}% <--- 必须加 %
    \hspace{1mm}% <--- 必须加 %
    % 第三张图
    \begin{subfigure}[b]{0.29\textwidth}
        \centering
        \includegraphics[width=\linewidth]{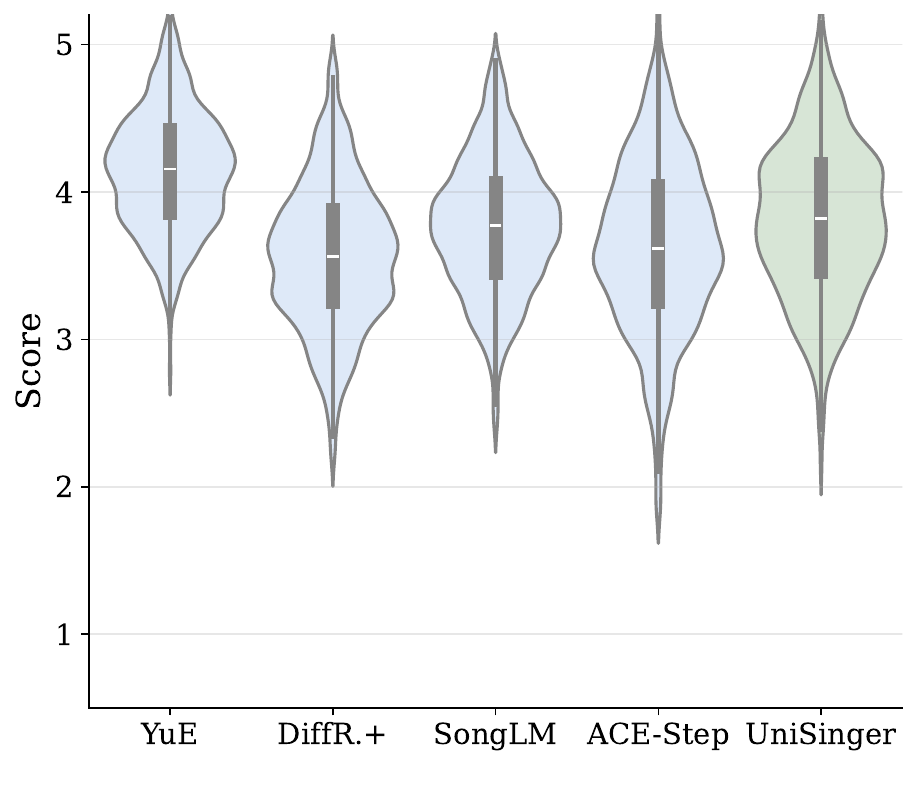}
        \vspace{-4ex}
        \caption{Quality}
        \vspace{-1ex}
        \label{fig:sub3}
    \end{subfigure}
    
    \caption{Comparison of subjective metrics. (a) shows Intelligibility, (b) demonstrates Similarity, and (c) presents the MOS Score.}
    \vspace{-4ex}
    \label{fig:three_comparison}
\end{figure*}

\vspace{-2ex}
\subsection{Curriculum Learning with Task-Specific Masking}\label{sec:masking}
\vspace{-1ex}

To mitigate multi-task optimization conflicts, we design a four-stage progressive curriculum learning strategy, as shown in the right pannel of Figure~\ref{fig:placeholder}. This strategy is enabled by a task-specific modality masking mechanism $\mathcal{M}$, formulated as:
\vspace{-1ex}
\begin{equation}
\label{eq:mask}
\begin{split}
    \tilde{c}_m &= \mathcal{M}(c_m, z_{\text{task}}) \\
    &= \mathbb{I}(m \in \mathcal{S}_z) \cdot c_m + (1 - \mathbb{I}(m \in \mathcal{S}_z)) \cdot \varnothing_m.
\end{split}
\end{equation}
% \vspace{-2ex}
\noindent where $c_m$ and $\tilde{c}_m$ denote the input and masked embeddings for modality $m$, respectively. $\mathbb{I}(\cdot)$ is the indicator function, and $\varnothing_m$ represents a learnable null token for the dropped modality. $\mathcal{S}_z$ denotes the set of active modalities determined by the task index $z{\text{task}}$. By progressively manipulating $\mathcal{S}_z$, we guide the model through the following stages.

\noindent\textbf{Stage 0: General Song Generation.}
We condition the model exclusively on text instructions and phonemes, denoted as $\mathcal{S}_0 = \{c_{txt}, c_{pho}\}$, enabling model to learn to map textual inputs to global musical attributes (e.g., genre, rhythm) and coarse-grained vocal timbres (e.g., gender, age), establishing foundational song generation capabilities..

\noindent\textbf{Stage 1: General SVC.} 
For SVC, the model is conditioned solely on semantic embeddings and speaker embeddings, represented as $\mathcal{S}_1 = \{c_{sem}, c_{spk}\}$. This creates an information bottleneck that compels the model to rely exclusively on $c_{\text{spk}}$ for speaker timbre reconstruction and establishing a robust unified speaker space.

\noindent\textbf{Stage 2: Speaker Cloning Song Generation.} 
We integrate speaker embeddings into song generation task with the input configuration $\mathcal{S}_2 = \{c_{txt}, c_{pho}, c_{spk}\}$. By combining vocal identity ($c_{\text{spk}}$), linguistic content ($c_{\text{txt}}$) and melodic information ($c_{\text{pho}}$), the model achieves high-fidelity zero-shot speaker cloning within song generation paradigm.

\noindent\textbf{Stage 3: Accompaniment Co-Generation SVC.} 
Finally, we extend SVC by introducing text instructions, formulated as $\mathcal{S}_3 = \{c_{\text{txt}}, c_{\text{sem}}, c_{\text{spk}}\}$. This compels the model to disentangle information streams: extracting linguistic content from semantic embeddings ($c_{\text{sem}}$), fine-grained vocal timbre from speaker embeddings ($c_{\text{spk}}$), and accompaniment attributes from text instructions ($c_{\text{txt}}$). Leveraging generative priors from the initial stages, the model synthesizes accompaniment harmoniously coordinated with the vocals.

\vspace{-1ex}
\subsection{Cross-Task Speaker Embedding Space}\label{sec:spk}
% \vspace{-1ex}
To ensure consistent vocal identity preservation across tasks, we construct a Cross-task Speaker Embedding Space.

\noindent\textbf{Speaker Representation via SVC.} 
To ensure cross-task vocal control, we use CAM++ feature as the speaker condition in SVC task. By training the model to reconstruct the target vocal timbre using only semantic and CAM++ feature, we build a unified cross-task speaker embedding space for the following tasks. 

\noindent\textbf{Cross-Task Representation Transfer.} 
Subsequently, this purified embedding is introduced into the song generation task as a potent acoustic prior. The model establishes a hierarchical control mechanism: utilizing text instructions to govern global musical styles, while employing the speaker embedding to precisely control fine-grained vocal attributes.

\vspace{-1ex}
\subsection{MM-DiT Backbone}
\vspace{-1ex}

As shown in Figure~\ref{fig:placeholder}(a), our framework employs an MM-DiT designed for flow matching, comprising a bottom stack of $N_1$ Joint DiT Layers and a top stack of $N_2$ Single DiT Layers.

\noindent\textbf{Joint DiT Layers.} 
The bottom $N_1$ layers are designed for cross-modal interaction. We concatenate the Queries, Keys, and Values from both text and audio modalities for joint attention processing. The output is then split back into original streams to align linguistic content with acoustic features.

\noindent\textbf{Single DiT Layers.} 
The top $N_2$ layers focus on refined audio modeling. Here, the text modality is discarded, and the operation reduces to self-attention on audio latents to polish fine-grained acoustic details.

During training, the model receives a timestep $t$ and learns a conditional velocity field $v_\theta$ to transform Gaussian noise into the target VAE latent. The objective is defined as:
\vspace{-1ex}
\begin{equation}
    \mathcal{L}_{\text{CFM}} = \mathbb{E}_{t, x_t} \left\lVert v_\theta(t, C_{\text{cond}}, x_t) - u(t, x_t) \right\rVert^2,
\end{equation}
% \vspace{-2ex}
where $u(t, x_t)$ is the optimal flow path and $C_{\text{cond}}$ denotes the condition input derived from the multi-modal processing module. During inference, an ODE solver integrates this field to generate the target latent, which is then synthesized into waveform by the Mel Decoder.

\vspace{-2ex}
\section{Experiments}
\subsection{Data and Training Details}
\vspace{-1ex}

% To construct our datasets, we collected 30k hours of in-the-wild songs standardized to 44.1kHz. As shown in Figure~\ref{fig:data}, the pipeline begins with rigorous preprocessing, where audio is filtered by Signal-to-Noise Ratio (SNR) ~\cite{johnson2006signal}, segmented into 15 to 30s clips via Voice Activity Detection (VAD) ~\cite{sohn1999statistical}, and processed with \textit{Hybrid Transformer Demucs} ~\cite{rouard2023hybrid} to isolate clean vocals for SVC tasks. Transcripts are generated through a voting mechanism leveraging Whisper-Large-v3~\cite{cao2012whisper}, SenseVoice~\footnote{~\url{https://github.com/FunAudioLLM/SenseVoice}}, and Qwen2.5-Omni~\cite{xu2025qwen2}. Furthermore, we utilize internal labeling model to perform fine-grained annotating, covering genre, rhythm, instrumentation, emotion, and singer demographics (age and gender). Finally, these annotations are merged with the transcripts to form unified dense captions, from which \textit{Qwen2.5-7B} ~\cite{hui2024qwen2} extracts the corresponding instruct embeddings.
We collected 30k hours of in-the-wild songs standardized to 44.1kHz. The preprocessing pipeline follows a standard flow: audio is filtered by SNR~\cite{johnson2006signal}, segmented via VAD~\cite{sohn1999statistical}, and separated using Hybrid Transformer Demucs~\cite{rouard2023hybrid} to obtain clean vocals. Transcripts are generated via a voting mechanism across Whisper-Large-v3~\cite{cao2012whisper} and Qwen2.5-Omni~\cite{xu2025qwen2}. Unified dense captions, incorporating fine-grained annotations (e.g., genre, rhythm, demographics), are processed by Qwen2.5-7B~\cite{hui2024qwen2} to extract instruction embeddings.

% For the training dataset, we utilize 20k hours from this processed corpus. During the final multi-task joint training stage, we incorporate an additional 5,000-hour high-quality internal dataset of accompanied singing, processed identically. From this internal corpus, we curate our evaluation dataset, consisting of 500 clips (15-20s) that strictly balance speaker diversity (5 males, 5 females) and linguistic distribution (Chinese and English).

% Our Model comprises 1.54 billion parameters with a flow matching feedforward dimension of 1024, including 14 joint DiT layers and 6 single DiT layers. Training is conducted on 16 NVIDIA Tesla A800 80GB GPUs with a batch size of 8 per GPU, using the Adam optimizer ~\cite{kingma2014adam} with an initial learning rate of 1e-4. 

We train the 1.54B parameter model (14 joint /6 single DiT layers, 1024 dff) using 20k hours of processed songs. For multi-task joint training, an additional 5k-hour internal dataset of accompanied singing is incorporated. The evaluation set comprises 500 balanced clips (15-20s) from the internal corpus, covering gender (5M/5F) and language (Chinese/English). Training uses Adam ($lr=1e-4$) on 16 NVIDIA A800 GPUs with a batch size of 8 per GPU.

\begin{table*}[t]
    \centering
    \footnotesize 
    \renewcommand{\arraystretch}{1.1}
    \setlength{\tabcolsep}{0pt}
    \caption{Evaluation Results of SVC. The best results are in bold, and second-best results are underlined.}
    \vspace{-2ex}
    \label{tab:svc_comparison}
    
    \begin{tabular*}{\textwidth}{l @{\extracolsep{\fill}} ccccccc}
        \toprule
        
        \multirow{2}{*}{\textbf{Model}} & 
        \multicolumn{3}{c}{\textbf{Objective Evaluation}} &
        \multicolumn{4}{c}{\textbf{Subjective Evaluation (MOS)}$\uparrow$} \\
        
        \cmidrule(lr){2-4}  \cmidrule(lr){5-8}
        
        & \textbf{PER}$\downarrow$ & \textbf{Spk-Sim}$\uparrow$ & \textbf{FPC}$\uparrow$ 
        & \textbf{Intelligibility} & \textbf{Similarity}  & \textbf{Quality} & \textbf{Harmony}\\
        
        \midrule
        
        HQ-SVC ~\cite{bai2025hq}  & 0.187 & 0.627 & \textbf{0.801} & 3.787 ± 0.124 & 3.678 ± 0.122 & \textbf{4.015 ± 0.124} & 3.412 ± 0.174 \\
        
        NeuCoSVC ~\cite{sha2024neural} & 0.243 & 0.573 & 0.612 & 3.824 ± 0.201 & \textbf{3.765 ± 0.114} & \underline{3.899 ± 0.112}  & 3.237 ± 0.188 \\
        
        So-VITS-SVC~\footnote{\url{https://github.com/svc-develop-team/so-vits-svc}} & \underline{0.154} & \underline{0.700} & 0.743 & \underline{3.910 ± 0.135} & 3.733 ± 0.231 & 3.721 ± 0.201 & 3.522 ± 0.218 \\
        \midrule

        UniSinger\_SVC  & \textbf{0.151} & \textbf{0.712} & \underline{0.771} & \textbf{3.912 ± 0.154} & \underline{3.758 ± 0.103} & 3.825 ± 0.107 & \underline{3.764 ± 0.138} \\

        UniSinger\_SVC\_BGM & 0.167 & 0.687 & 0.655 & 3.785 ± 0.210   & 3.612 ± 0.123 & 3.771 ± 0.143 & \textbf{3.891 ± 0.109} \\
        
        \bottomrule
    \end{tabular*}
\end{table*}
\vspace{-2ex}

% \subsection{Training Details}
% Our Model comprises 1.54 billion parameters with a flow matching feedforward dimension of 1024. The architecture includes 14 joint diffusion transformer layers and 6 single diffusion transformer layers, incorporating RoPE positional encoding ~\cite{lazos2005rope}. Training is conducted on 16 NVIDIA Tesla A800 80GB GPUs with a batch size of 8 per GPU, using the Adam optimizer ~\cite{kingma2014adam} with an initial learning rate of 1e-4. 

% During multi-stage progressive training, to prevent catastrophic forgetting during task transitions, we maintain a 30\% data replay buffer from previous stages, culminating in a final joint training phase to consolidate unified capabilities. 

\vspace{-1ex}
\subsection{Evaluation Metrics}
\vspace{-1ex}

% \textbf{Objective Evaluation} 
% We measure Phoneme Error Rate (PER) using FireRedASR~\cite{xu2025fireredasr} and Speaker Similarity (Spk-Sim) using a WavLM-based model~\cite{chen2022wavlm}. Semantic alignment and musical quality are evaluated using CLaMP 3~\cite{wu2025clamp} and SongEval~\cite{yao2025songeval}, respectively. For SVC tasks, we additionally report Fréchet Audio Distance (FAD)~\cite{kilgour2018fr} for overall fidelity and $F_0$ Pearson Correlation (FPC) for pitch accuracy.

% \textbf{Subjective Evaluation}
% We conduct Mean Opinion Score (MOS) tests with 30 listeners (20 music experts, 10 non-experts) rating samples on a 1–5 scale for musicality, audio quality, and intelligibility. We also introduce a Harmony metric to assess vocal-BGM rhythmic coordination. To ensure fair comparison, we employ SingSong to synthesize BGM for baseline SVC models that generate vocals only.

We evaluate performance using objective metrics: Phoneme Error Rate (PER) using FireRedASR~\cite{xu2025fireredasr}, Speaker Similarity (Spk-Sim) using a WavLM-based model~\cite{chen2022wavlm}, CLaMP 3~\cite{wu2025clamp} for semantics, and SongEval ~\cite{yao2025songeval} for music quality. SVC tasks also include Fréchet Audio Distance (FAD)~\cite{kilgour2018fr} and $F_0$ Pearson Correlation (FPC). Subjective Mean Opinion Score (MOS) tests were conducted by 30 listeners across musicality, quality, and intelligibility, plus a Harmony metric for vocal-BGM coordination. For fair comparison, baseline SVC models use SingSong to generate BGM.

\vspace{-1ex}
\subsection{Speaker Cloning Song Generation}
\vspace{-1ex}

% 对比模型
We compare UniSinger against four representative state-of-the-art models: \textit{SongLM}, a language-model-based framework for text-to-song generation; \textit{YuE}, a large-scale model capable of generating full songs with zero-shot speaker cloning; \textit{ACE-Step}, a discrete token-based generation approach; and \textit{DiffRhythm+}, a diffusion-based model supporting strict lyric alignment and speaker similarity. And we use the UniSinger\_Song to represent the song generation task model.

% \vspace{-1ex}
% \subsubsection{Objective Results}
% \vspace{-1ex}

% The objective results are summarized in Table~\ref{tab:song_gen_results}. For intelligibility, our model achieves the lowest PER of 19.61\%, surpassing the strong diffusion baseline DiffRhythm$^+$ of 20.72\%. We attribute this precision to the joint training with the SVC task, which demands frame-level articulation reconstruction, enhancing the model's sensitivity to phonemic details. Second, UniSinger attains the highest Speaker Similarity of 68.85\%. This score establishes far exceeds standard text-to-music models like SongLM. This result validates the effectiveness of our Cross-task Speaker Space in transferring vocal identity from the SVC task to song generation. Finally, in terms of musicality, UniSinger achieves top performance in most SongEval metrics. Specifically, it leads in Coherence and Melody, demonstrating that unifying tasks facilitates the generation of structurally integrity and melodic richness. While YuE shows advantages in CLaMP 3, UniSinger delivers a superior overall balance, offering higher acoustic fidelity and speaker control while maintaining competitive musical structure.

Table~\ref{tab:song_gen_results} summarizes the objective results. Regarding intelligibility, our model achieves the lowest PER of 19.61\%, surpassing the strong baseline DiffRhythm+ of 20.72\%. We attribute this precision to joint training with SVC, which demands frame-level reconstruction and enhances phonemic sensitivity. Second, UniSinger attains the highest Speaker Similarity of 68.85\%, far exceeding standard text-to-music models like SongLM. This validates the effectiveness of our Cross-task Speaker Space in transferring vocal identity. Finally, regarding musicality, UniSinger leads in most SongEval metrics, specifically Coherence and Melody, demonstrating that unifying tasks facilitates structural integrity and melodic richness. While YuE shows advantages in CLaMP 3, UniSinger delivers a superior balance of acoustic fidelity, speaker control, and competitive musical structure.

% \vspace{-1ex}
% \subsubsection{Subjective Results}
% \vspace{-1ex}

% The subjective results are shown in the violin plots of Figure~\ref{fig:three_comparison}. 
% It is observed that the large-scale model YuE achieves the highest median scores in Similarity and Quality. We attribute this advantage to YuE's massive scale of 3B parameters and 650k hours training data, which endow it with capabilities in semantic modeling and acoustic diversity. Nevertheless, UniSinger remains highly competitive, demonstrating particular strength in linguistic clarity.  As shown in Figure~\ref{fig:three_comparison}(a), UniSinger exhibits a score distribution heavily skewed towards the upper range for Intelligibility, matching the performance of YuE and visibly outperforming DiffRhythm+. This subjective preference aligns perfectly with the objective PER results in Table~\ref{tab:song_gen_results}, confirming that our unified training strategy effectively refines pronunciation without requiring billion-scale parameters. Furthermore, in terms of Similarity and Quality, UniSinger maintains a robust distribution significantly superior to SongLM and ACE-Step, and rivals the strong baseline DiffRhythm+. This highlights that UniSinger delivers high-fidelity generation while offering more precise vocal control and resource efficiency.

Figure~\ref{fig:three_comparison} presents the subjective results. The large-scale model YuE achieves the highest median scores in Similarity and Quality, which we attribute to its massive 3B parameters and 650k hours of training data. Nevertheless, UniSinger remains highly competitive, demonstrating particular strength in linguistic clarity. As shown in Figure~\ref{fig:three_comparison}(a), UniSinger matches the Intelligibility performance of YuE and visibly outperforms DiffRhythm+. This aligns with the objective PER results in Table~\ref{tab:song_gen_results}, confirming that our strategy effectively refines pronunciation without requiring massive parameters. Furthermore, regarding Similarity and Quality, UniSinger significantly outperforms SongLM and ACE-Step and rivals the strong baseline DiffRhythm+. This highlights that UniSinger delivers high-fidelity generation with precise vocal control.

\vspace{-2ex}
\subsection{Singing Voice Conversion with Accompaniment}
\vspace{-1ex}

We compare UniSinger with three SVC models: \textit{NeuCoSVC}, a neural concatenative framework employing self-supervised feature matching; \textit{So-VITS-SVC}, a VITS-based model utilizing a SoftVC encoder for content-timbre disentanglement; \textit{HQ-SVC}, a zero-shot framework leveraging decoupled audio codecs and DDSP-diffusion for low-resource scenarios. Additionally, we use UniSinger\_SVC and UniSinger\_SVC\_BGM to denote the model variants for pure SVC tasks and for SVC tasks with accompaniment co-generation.

% \subsubsection{Objective Results}
% As summarized in Table~\ref{tab:svc_comparison}, UniSinger demonstrates superior performance in content preservation and speaker similarity. Specifically, it achieves the lowest PER of 0.151, surpassing the strong baseline So-VITS-SVC at 0.154, thereby proving its robustness in linguistic reconstruction. Additionally, UniSinger obtains the highest Spk-Sim score of 0.712, significantly outperforming all baselines. Regarding pitch accuracy, while HQ-SVC achieves the top FPC of 0.801, UniSinger remains highly competitive as the runner-up with a score of 0.771. For the accompaniment-coordinated variant UniSinger\_BGM, integrating background music introduces expected generative interference and a slight metric decline. However, despite the complex objective of simultaneous generation, UniSinger\_BGM maintains remarkable robustness. It achieves a PER of 0.167 and a Spk-Sim of 0.687, surprisingly outperforming dedicated dry-vocal baselines such as HQ-SVC and NeuCoSVC. This confirms our architecture effectively manages complex acoustic interactions while delivering vocal fidelity that rivals traditional single-task systems.

As summarized in Table~\ref{tab:svc_comparison}, UniSinger demonstrates superior performance in content preservation and speaker similarity. Specifically, it achieves the lowest PER of 0.151, surpassing the strong baseline So-VITS-SVC at 0.154. Additionally, UniSinger obtains the highest Spk-Sim score of 0.712, significantly outperforming all baselines. Regarding pitch accuracy, UniSinger remains highly competitive with an FPC of 0.771, ranking second only to HQ-SVC. For the accompaniment variant UniSinger\_BGM, integrating background music causes a slight expected metric decline. However, it maintains remarkable robustness with a PER of 0.167 and Spk-Sim of 0.687. Surprisingly, these scores outperform dedicated dry-vocal baselines like HQ-SVC and NeuCoSVC. This confirms that our architecture effectively manages complex acoustic interactions while delivering vocal fidelity that rivals single-task systems.

For the subjective results, UniSinger achieves the highest Intelligibility of 3.912, slightly surpassing So-VITS-SVC at 3.910, while maintaining competitive Speaker Similarity. Although Audio Quality trails HQ-SVC due to minor artifacts from in-the-wild training data, the overall performance remains robust. For the accompaniment variant UniSinger\_BGM, while joint modeling introduces a slight expected degradation in base metrics, it establishes a decisive advantage in Harmony. With a score of 3.891, it significantly outperforms cascaded baselines like HQ-SVC and NeuCoSVC. This confirms that our end-to-end framework produces superior acoustic coherence and rhythmic synergy compared to traditional multi-stage synthesis.

\vspace{-1ex}
\begin{table}[t]
    \centering
    \small
    \setlength{\tabcolsep}{0pt}
    \vspace{-3ex}
    \caption{Ablation study on training strategies. }
    \vspace{-1ex}
    \label{tab:ablation}
    
    \begin{tabular*}{\linewidth}{l @{\extracolsep{\fill}} ccc}
        \toprule
        \textbf{Settings} & \textbf{PER}$\downarrow$ & \textbf{Spk-Sim}$\uparrow$ & \textbf{Harmony}$\uparrow$ \\
        
        \midrule

        \textbf{UniSinger\_Song} 
        & 19.61\% & 68.85\% & - \\
        
        \textit{w/o Task Masking} 
        & 25.83\% & 60.99\% & - \\ 

        \textit{w/o SVC Stage} 
        & 22.69\%  & 53.24\% & - \\ 

        \textit{w/o Speaker Broadcast} 
        & 18.23\% & 50.24\% & - \\

        \midrule

        \textbf{UniSinger\_SVC\_BGM} 
        & 15.18\% & 68.43\% & 3.98 \\
        
        \textit{w/o Song Gen Stage} 
        & 16.23\% & 68.19\%  & 2.83 \\

        % \midrule 

        \bottomrule
    \end{tabular*}
\end{table}

% \begin{table}[t]
%     \centering
%     \small
%     \setlength{\tabcolsep}{0pt}
%     \caption{Ablation study on training strategies. }
%     \label{tab:ablation}
    
%     \begin{tabular*}{\linewidth}{l @{\extracolsep{\fill}} ccc}
%         \toprule
%         \textbf{Settings} & \textbf{PER}$\downarrow$ & \textbf{Spk-Sim}$\uparrow$ & \textbf{Harmony}$\uparrow$ \\
        
%         \midrule
    
%         \textit{w/o Task Masking} 
%         & 25.83\% & 60.99\% & - \\ 
%         \midrule

%         % \textit{Repl. Broadcast w/ AdaLN} 
%         % & 18.23\% & 50.24\% & - \\ 
        
%         \textit{w/o SVC Stage} 
%         & 22.69\%  & 53.24\% & - \\ 

%         \textit{w/o Song Gen Stage} 
%         & 16.23\% & 68.19\%  & 2.83 \\ 

%         \textbf{Ours} 
%         & 19.61\% & 68.85\% & - \\

%         \textbf{Ours} 
%         & 19.61\% & 68.85\% & - \\
%         \midrule

%         \textit{Repl. Broadcast w/ AdaLN} 
%         & 18.23\% & 50.24\% & - \\ 

%         \textbf{Ours} 
%         & 19.61\% & 68.85\% & - \\

%         % \textbf{UniSinger\_SVC\_BGM} 
%         % & 15.18\% & 68.43\% & 3.98 \\
    
%         % \textit{w/o Song Gen Stage} 
%         % & 16.23\% & 68.19\%  & 2.83 \\ 
       
%         % \midrule 

%         \bottomrule
%     \end{tabular*}
% \end{table}
% \vspace{-1ex}
\subsection{Ablation Study}
\vspace{-1ex}
% We conduct ablation studies to validate the effectiveness of our Task-Specific Modality Masking, Speaker Embedding Injection Strategy, and Multi-Stage Curriculum Learning. 
% The ablation study results are summarized in Table ~\ref{tab:ablation}. First, we remove Task-Specific Masking, naively concating all input modalities and inputing into model across all training stages. Using Timbre-Controlled Song Generation Task as a case study, we observe significant degradation in both PER and Spk-Sim. This confirms that our masking strategy is essential for resolving multi-task optimization conflicts.Next, we change the speaker embedding injection strategy from broadcasting to direct global injection via AdaLN. This alteration reveals a trade-off: While this alteration yields a lower PER, it simultaneously causes a significant drop in Spk-Sim. This suggests that while AdaLN favors linguistic clarity, it compromises fine-grained speaker cloning capability, thereby confirming that our broadcasting mechanism has a better balance in terms of intelligibility and speaker similarity. Finally, we evaluate the contribution of multi-stage training strategy. Removing the SVC Stage causes a significant deterioration in Spk-Sim and a slight degradation in PER for Song Generation. Similarly, eliminating the Song Generation Stage leads to a marked decline in the Harmony metric for SVC. These findings confirm that every stage of our progressive curriculum is indispensable for achieving unified capabilities.
Table~\ref{tab:ablation} summarizes the ablation study. First, removing Task-Specific Masking by naively concatenating all input modalities causes significant degradation in both PER and Spk-Sim for song generation. This proves our masking strategy is vital for resolving multi-task optimization conflicts. Next, replacing our speaker embedding broadcasting with direct global injection via AdaLN reveals a trade-off. Although AdaLN improves PER, it severely compromises Spk-Sim, confirming our broadcasting mechanism provides a superior balance between linguistic clarity and fine-grained speaker cloning. Finally, evaluating the multi-stage training reveals its necessity. Removing the SVC stage significantly degrades Spk-Sim in song generation, while eliminating the song generation stage causes a marked decline in the SVC Harmony metric. Therefore, every stage of our progressive curriculum is indispensable for unified performance.

% 首先,我们去掉task-specific Masking,每个训练阶段都将input模态全部输入进模型,模型需要自己去学习需要使用哪个输入模态,我们以Timbre-Controllable Song Generation任务为例,测试模型去掉task-specific Masking后,经过四阶段训练的结果,发现：
% 接着,我们将spk embedding的注入方式从Broadcast改为AdaLN方式直接注入全局表征,同样以以Timbre-Controllable Song Generation任务为例,发现spk-sim有了明显下降,推测是因为全局表征的控制能力很弱.....
% 最后,我们分别消融SVC训练阶段和Song Gen训练阶段,发现,消融SVC阶段只训练Song Gen阶段会导致sim的明显下降。

% 最后,消融Song Gen阶段只训练SVC阶段会导致harmony的下降
% 最后,我们对多阶段训练做消融, Remove the SVC Training Stage,测试song generation的性能,发现SPk-Sim有了明显的劣化,同时PER也有了轻微的下降。同样,remove song generation training stage,发现对应的SVC的伴奏和谐度指标也有明显的下降。说明了多阶段训练每一个阶段都是不可或缺的。
% 

\vspace{-2ex}
\section{Conclusion}
\vspace{-1ex}

We present UniSinger, the first end-to-end framework that unifies the historically isolated paradigms of song generation and SVC. By constructing a cross-task speaker embedding space, we successfully bridge the gap between semantic understanding and acoustic control. Furthermore, our curriculum learning strategy, driven by task-specific modality masking, effectively resolves the optimization conflicts in multi-task training. This approach not only enables the integration of both tasks, but also facilitates reciprocal enhancement. By establishing vocal-accompaniment synergy, our framework pioneers the new capability of accompaniment co-generation SVC, offering an intelligent solution for AI music production.

\section{Generative AI Use Disclosure}
During the preparation of this work, the authors used Gemini to assist with refining the grammatical structure. The authors reviewed and edited the output and take full responsibility for the content of the publication. 

\bibliographystyle{IEEEtran}
\bibliography{newbib}

\end{document}